\documentclass[prb,twocolumn,superscriptaddress,floats]{revtex4}
\usepackage{graphicx}
\usepackage{amsmath}
\usepackage{booktabs}
\usepackage{dcolumn}
\usepackage{textcomp}
\usepackage{chemsym}

\usepackage{rotating}

\newcolumntype{d}{D{.}{.}{-1}}

\newcommand{\knif}{K$_2$NiF$_4$}

\newcommand{\lsco}{La$_{2-x}$Sr$_{x}$CuO$_4$}
\newcommand{\lsmo}{La$_{1-x}$Sr$_{1+x}$MnO$_4$}
\newcommand{\lsmoacht}{La$_{0.875}$Sr$_{1.125}$MnO$_4$}
\newcommand{\lsmoviert}{La$_{0.75}$Sr$_{1.25}$MnO$_4$}
\newcommand{\remoper}{RE$_{1-x}$(Sr/Ca)$_{x}$MnO$_3$}
\newcommand{\lmo}{LaMnO$_3$}
\newcommand{\lsmoef}{La$_{0.5}$Sr$_{1.5}$MnO$_4$}
\newcommand{\lsmoee}{La$_{1.0}$Sr$_{1.0}$MnO$_4$}

\newcommand{\lsmodzs}{La$_{2-2x}$Sr$_{1+2x}$Mn$_2$O$_7$}

\newcommand {\vq}{$\mathbf{q}$}

\newcommand {\mB}{$\mu _B$}

\newlength{\figwidth}

\divide\figwidth by 2

\begin{document}

\advance\vsize by 2 cm

\title{Crystal and magnetic structure of La$_{1-x}$Sr$_{1+x}$MnO$_4$ : role of the orbital degree of freedom}

\author{D. Senff}
\affiliation{II. Physikalisches Institut, Universit\"at zu K\"oln, Z\"ulpicher Str. 77, D-50937 K\"oln, Germany}

\author{P. Reutler}
\affiliation{II. Physikalisches Institut, RWTH Aachen, Huykensweg
D-52056 Aachen, Germany} \affiliation{Laboratoire de
Physico-Chimie de l'Etat Solide, Universit\'e Paris Sud, 91405
Orsay Cedex, France}

\author{M. Braden}
\email{braden@ph2.uni-koeln.de}%
\affiliation{II. Physikalisches Institut, Universit\"at zu K\"oln, Z\"ulpicher Str. 77, D-50937 K\"oln, Germany}

\author{O. Friedt}
\affiliation{II. Physikalisches Institut, Universit\"at zu K\"oln, Z\"ulpicher Str. 77, D-50937 K\"oln, Germany}

\author{D. Bruns}
\affiliation{II. Physikalisches Institut, Universit\"at zu K\"oln, Z\"ulpicher Str. 77, D-50937 K\"oln, Germany}
\affiliation{II. Physikalisches Institut, RWTH Aachen, Huykensweg D-52056 Aachen, Germany}

\author{A. Cousson}
\affiliation{ Laboratoire L\'eon Brillouin, C.E.A./C.N.R.S., F-91191 Gif-sur-Yvette Cedex, France}

\author{F. Bour\'ee}
\affiliation{ Laboratoire L\'eon Brillouin, C.E.A./C.N.R.S., F-91191 Gif-sur-Yvette Cedex, France}

\author{M. Merz}
\affiliation{Institut f\"ur Kristallographie, RWTH Aachen, D-52056
Aachen, Germany}

\author{B. B\"uchner}
\affiliation{IFW-Dresden, Helmholtzstrasse 20, D-01069 Dresden,
Germany}

\author{A. Revcolevschi}
\affiliation{Laboratoire de Physico-Chimie de l'Etat Solide,
Universit\'e Paris Sud, 91405 Orsay Cedex, France}

\date{\today, \textbf{preprint}}

\pacs{PACS numbers: 78.70.Nx, 75.40.Gb, 74.70.-b}

\begin{abstract}

The crystal and magnetic structure of La$_{1-x}$Sr$_{1+x}$MnO$_4$ (0$\,\leq\,$x$\,\leq\,$0.7) has been studied by
diffraction techniques and high resolution capacitance dilatometry. There is no evidence for a structural phase
transition like those found in isostructural cuprates or nickelates, but there are significant structural changes
induced by the variation of temperature and doping which we attribute to a rearrangement of the orbital
occupation.

\end{abstract}

\maketitle

\section{Introduction}

Manganates have drawn a large interest due to the observation of a colossal magneto-resistivity in \remoper{} and
in \lsmodzs{}.\cite{mang-gen, magres-327} The magnetoresistivity appears to result from the competition between
the ferromagnetic metallic state and the ordering of charge, orbital and magnetic degrees of freedom. The single
layer materials \lsmo , of the \knif -structure type, have been much less studied, because in these compounds
high magnetic fields of the order of 30\ T are needed to induce a large magneto-resistivity.\cite{mag-214} The
high value of the transition field, however, indicates that the complex ordered state is particularly stable in
the single layer manganates. Many techniques have been applied to characterize the order in the half-doped
layered compound \lsmoef.\cite{sternlieb,murakami,larochelle,wilkins,dhesi}

The composition \lsmoee{} corresponds to \lmo{} in the perovskite series, since all Mn-ions are three-valent.
Like the perovskite \lmo , \lsmoee{} exhibits an antiferromagnetic order. Kawano et al. report a small ordered
moment of only 0.8 \mB{} \cite{kawano} compared to the value of 3.87(3) observed in \lmo.\cite{moussa} A more
recent study on \lsmoee{} finds a larger moment of 3.3\mB.\cite{larger-m} Through the substitution of La by
divalent Sr, part of the Mn is oxidized, which may suggest physics similar to that of the perovskite materials
based on the Zener double exchange mechanism. However, in the \lsmo \ series there is no metallic ferromagnetic
phase at ambient conditions. Instead, the charges, the orbital degrees of freedom and the spins order in closely
coupled patterns. In this sense the \lsmo{} phase diagram resembles that of Pr$_{1-x}$Ca$_x$MnO$_3$ where there
is no metallic phase at zero magnetic field neither. Upon increasing the Sr content in \lsmo \ the commensurate
antiferromagnetic order disappears near x=0.15 \cite{bao,moritomo} and charge order appears for x larger
than 0.4.\cite{moritomo,larochelle} For the intermediate concentrations spin-glass behavior is
observed.\cite{moritomo}

Concerning the crystal structure, so far only the lattice parameters were determined as a function of
doping.\cite{bao} Upon introduction of Sr, the $c$ parameter decreases, whereas the in-plane parameter $a$
increases. We have performed a systematic study of the crystal and magnetic structure of \lsmo{} by neutron and
by x-ray diffraction techniques combined with measurements of the thermal expansion. We find strong evidence of
an orbital rearrangement, both as a function of temperature and as a function of doping.

\begin{figure}[tp]
\resizebox{0.9\figwidth}{!}{
\includegraphics*[angle=-0]{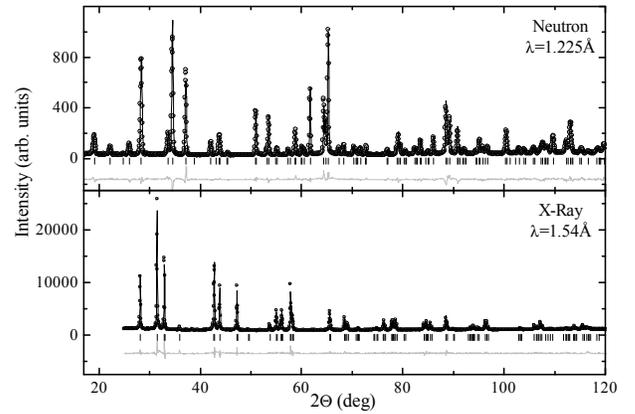}}
\caption{Typical diffraction patterns obtained for the layered
manganates : (top) a neutron powder diffraction pattern measured
on the 3T.2 diffractometer for \lsmoviert; (bottom) a x-ray
diffraction pattern determined on a D5000-diffractometer. Points
designate the intensity data, dark lines the profile fits and
fine lines the intensity differences. Vertical bars indicate the
the Bragg-positions.}\label{fig1}
\end{figure}

\section{Experimental}

Single crystals of \lsmo{} were grown as described elsewhere.\cite{reutler} Powder samples were obtained by
crushing single crystalline samples with a poor mosaic spread. Magnetic properties of the samples were reported
by Baumann et al..\cite{baumann} The crystal and magnetic structure of \lsmo{} was studied by powder and by
single crystal diffraction using neutrons and x-rays. All neutron experiments were performed on instruments of
the Laboratoire L\'{e}on Brillouin installed at the Orph\'ee reactor in Saclay. The crystal and magnetic
structure of single crystals of composition \lsmoee{} and \lsmoef{} were studied on the neutron four-circle
diffractometer 5C.2 at room temperature and at $\sim$20\ K. Powder neutron diffraction experiments on samples of
\lsmo{} with x=0.125, 0.25, 0.4 and 0.6 were performed on the high resolution neutron diffractometer 3T.2 using a
wavelength of 1.23 \AA . The temperature dependence of the lattice parameters was studied by powder x-ray
diffraction using Cu-K$_\alpha$ radiation and by measurements of the thermal expansion coefficients with a high
resolution capacitance dilatometer on single crystalline samples. All diffraction patterns were analyzed by the
Rietveld method using the Fullprof program.\cite{fullprof} Typical diffraction patterns, together with the
profile-fitted description, are shown in Fig. 1. The temperature dependence of magnetic superstructure peaks was
determined using the 4F triple-axis spectrometers.

\section{Magnetic and crystal structure of low-doped samples with commensurate antiferromagnetic order}

\lsmoee{} exhibits antiferromagnetic order with the magnetic moments aligned along the $c$ axis.\cite{kawano}
Using a small single crystal of $\sim$10mm$^3$ volume a set of 1100 Bragg reflection intensities was collected on
the four-circle diffractometer 5C.2 in respect with the I4/mmm lattice (dimension
$3.8\times3.8\times12\text{\AA}^3$). At room temperature, no evidence for superstructure reflections was found.
At low temperature, superstructure intensities were found at positions ($h/2$ $k/2$ l) in respect with the I4/mmm
lattice which agree with the antiferromagnetic order reported previously.\cite{kawano,larger-m,moritomo} A set of
152 magnetic Bragg reflection intensities and 965 fundamental intensities was recorded at 20\ K. Due to the high
crystal quality, extinction effects turned out to be important in this experiment and that on the composition
\lsmoef, see below. Therefore, we have corrected the data for extinction using the Becker-Coppens formalism for
an anisotropic mosaic spread. It is important to take into account the facts that the reliability of the
determination of the strong Bragg intensities is hampered by the extinction and that the multiple diffraction may
add intensity to weak reflections. Therefore, we have enhanced the statistical errors $\sigma_{stat}$ by :
\[
    \sigma ^2=\sigma^2_{stat} + Int.*c_{mult} + c_{const}
\]
using the observed intensity $Int$ and two additional constants $c_{mult}$ and $c_{const}$, following the
procedure described in detail for a similar problem.\cite{braden-cgo} In addition, reflections with an extinction
correction larger than 0.75 were excluded from the refinements. Using this procedure we obtain a satisfying
description of the Bragg reflections. The sets of 1037(871) reflections taken at room temperature (T=20\ K) are
described with reliability factors of $R_w(F^2)$=4.71\%(3.16\%) and $R_{unw.}(F^2)$=3.99\%(3.56\%). The
structural results are given in Table I.

\begin{table*}
    \begin{ruledtabular}
    \begin{tabular}{lcrrrrrrr}
                &                                  &      & x=0\footnotemark[1]    & x=0.125\footnotemark[2] & x=0.25\footnotemark[2] & x=0.4\footnotemark[2] & x=0.5\footnotemark[1] & x=0.6\footnotemark[2] \\[0.4ex] \hline\\[-1.6ex]
     a (\AA)    &                                  &   RT &     3.786 &      3.814 &      3.846 &      3.857 &       3.863&      3.857\\
                &                                  &   LT &     3.768 &      3.794 &      3.840 &      3.852 &       3.855&      3.857\\
     c (\AA)    &                                  &   RT &    13.163 &     12.938 &     12.676 &     12.548 &      12.421&     12.402\\
                &                                  &   LT &    13.195 &     12.985 &     12.651 &     12.524 &      12.397&     12.405\\
     V (\AA)$^3$&                                  &   RT &   188.676 &    188.204 &    187.500 &    186.670 &     185.356&    184.498\\
                &                                  &   LT &   187.348 &    186.879 &    186.588 &    185.855 &     184.257&    184.555\\[0.4ex]\hline\\[-1.6ex]
     Mn         & $U_{iso}$ $(10^{-4}\text{\AA}^2)$&   RT &    13(3)/68(3)\footnotemark[3]      &      15(6) &       18(6)&        5(4)& 19(2)/30(3)\footnotemark[3] & 28(5)\\
                &                                  &   LT &     9(3)/30(5)\footnotemark[3]      &       5(4) &       29(7)&          10\footnotemark[4] &  5(3)/ 5(3)\footnotemark[3] &  6(4)\\[1.2ex]
    La/Sr       &  z                               &   RT & 0.35598(2)& 0.35688(8) &  0.35815(9)& 0.35917(11)&  0.35816(3)& 0.35799(8)\\
                &                                  &   LT & 0.35564(2)& 0.35699(8) &  0.35779(9)& 0.35864(15)&  0.35814(4)& 0.35786(7)\\
                & $U_{iso}$ $(10^{-4}\text{\AA}^2)$&   RT &    46(8)/38(8)\footnotemark[3]      &      34(3) &       57(3)&       63(3)& 40(1)/30(1)\footnotemark[3] & 7(4)\\
                &                                  &   LT &    19(1)/17(2)\footnotemark[3]      &      18(3) &       41(3)&       10(4)&  5(3)/ 5(3)\footnotemark[3] & 6(2)\\[1.2ex]
     O(1)       & $U_{11}$ $(10^{-4}\text{\AA}^2)$ &   RT &      31(1)&      46(8) &       63(9)&      122(9)&       62(2)&      67(6)\\
                &                                  &   LT &      27(2)&      63(8) &       83(9)&       43(8)&       36(3)&      54(6)\\
                & $U_{22}$ $(10^{-4}\text{\AA}^2)$ &   RT &      71(1)&      46(7) &       75(8)&       65(8)&       69(2)&      47(6)\\
                &                                  &   LT &      45(2)&      24(8) &       87(8)&       97(8)&       68(3)&      66(6)\\
                & $U_{33}$ $(10^{-4}\text{\AA}^2)$ &   RT &     100(2)&      74(8) &      55(11)&      94(11)&       67(2)&      87(7)\\
                &                                  &   LT &      58(3)&      62(8) &      55(10)&       6(10)&       27(4)&      42(6)\\[1.2ex]
     O(2)       &  z                               &   RT & 0.17221(3)& 0.16900(11)& 0.16494(17)& 0.16386(15)&  0.16106(8)&0.15966(10)\\
                &                                  &   LT & 0.17270(4)& 0.17023(11)& 0.16449(15)& 0.16260(19)& 0.16138(10)& 0.15968(9)\\
                & $U_{11}$ $(10^{-4}\text{\AA}^2)$ &   RT &     179(2)&     191(7) &     167(15)&      153(7)&      117(2)&      93(4)\\
                &                                  &   LT &     131(2)&      86(6) &      131(7)&      134(9)&       70(2)&      63(4)\\
                & $U_{33}$ $(10^{-4}\text{\AA}^2)$ &   RT &      74(2)&     163(9) &     191(11)&     137(10)&       68(2)&      61(7)\\
                &                                  &   LT &      51(3)&     180(9) &     185(10)&       56(8)&       52(3)&      33(6)\\[0.4ex]\hline\\[-1.6ex]
    Bond length & $d_{\Mn-\O(1)}$ (\AA)            &   RT & 1.89335(1)& 1.90714(3) &  1.92329(4)&  1.92903(4)&  1.93164(3)& 1.92853(3)\\
                &                                  &   LT & 1.88403(8)& 1.89684(4) &  1.92018(4)&  1.92615(5)&  1.92763(6)& 1.92859(3)\\
                & $d_{\Mn-\O(2)}$ (\AA)            &   RT &  2.2668(4)& 2.1873(14) &  2.0915(21)&  2.0574(19)&   2.0005(9)& 1.9805(10)\\
                &                                  &   LT &  2.2788(5)& 2.2104(14) &  2.0810(19)&  2.0360(24)& 2.0006(13) & 1.9808(5)
    \end{tabular}
    \end{ruledtabular}
    \begin{flushleft}{\footnotesize
    \footnotemark[1]Refinement of single-crystal-data
    \footnotemark[2]Refinement of powder-crystal-data
    \footnotemark[3]The single-crystal data refinement allows one
    to determine anisotropic thermal parameters on the Mn and La/Sr sites, U11=U22 and U33.
    \footnotemark[4]This value was fixed during the refinement since it tended to negative values.
    }
    \end{flushleft}
    \vspace{-2.5ex}
    \caption{Results of the single crystal and powder neutron diffraction experiments on \lsmo{} for room (RT)
    and low temperature (LT).}
\end{table*}

In the \knif -structure of space group I4/mmm, there is an intrinsic difference between the antiferromagnetic
ordering of spins oriented along the $c$ direction or oriented along an in-plane direction. If the spins are
oriented within the plane, the phase of the order in  the neighboring plane leads to two symmetrically different
phases usually called La$_2$CuO$_4$ and La$_2$NiO$_4$-type. In contrast the different phases of ordering along
the $c$ direction yield two domains of the same symmetry if the ordered moment points along the $c$ direction.
The three-dimensional coupling in \lsmoee{} arises from the coupling with the next-nearest layer, which is
ferromagnetic and rather weak; the two magnetic domains are drawn in Fig. 2.

\begin{figure}[tp]
\resizebox{0.95\figwidth}{!}{
\includegraphics*{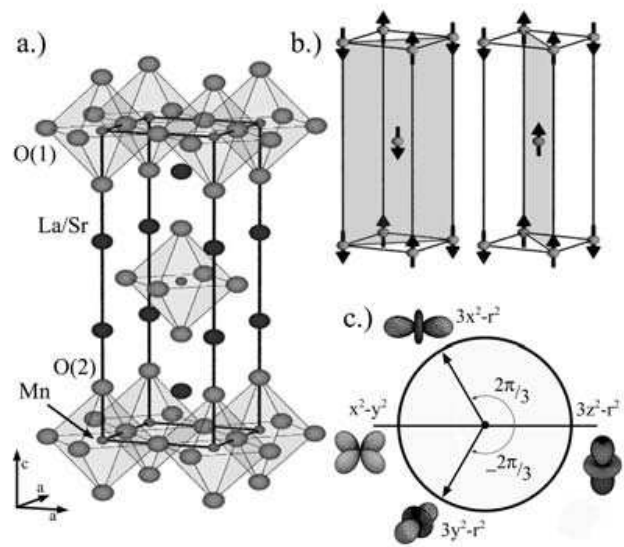}
} \caption{ a) Crystal structure of \lsmoee{} at room temperature; b) antiferromagnetic structure of \lsmoee ,
the two domains are shown (the ferromagnetic planes are shaded for the two domains); c) schematic drawing of the
$e_g$ orbital state in the pseudospin space. The angles $\theta=-{2\pi}/3, \pi, {2\pi}/3$ and $0$ correspond to
the $3y^2-r^2$, $x^2-y^2$, $3x^2-r^2$ and $3z^2-r^2$ orbitals, respectively.}\label{fig2}
\end{figure}

The collected magnetic Bragg intensities correspond to a superposition of contributions of both domains. The
refinement of the magnetic structure was performed with the Fullprof program using the magnetic form factor of
Mn$^{3+}$ and the crystal structure determined above. The two-domain structure was taken into account. We obtain
an ordered magnetic moment of 3.21(3)\mB{} in strong disagreement with a first report,\cite{kawano} but in good
agreement with more recent work.\cite{larger-m} The difference is most likely due to a poor sample quality of the
crystal studied first. Our sample exhibits the antiferromagnetic ordering at T$_N$=127\ K, in good agreement with
the study by Moritomo et al.\cite{moritomo} Attempts to refine an additional ordered moment component, aligned
perpendicular to the $c$  axis, did not yield a significant value. The ordered moment is still lower than what is
expected for a Mn$^{3+}$-ion without an orbital contribution, i.e. 4\mB , and it is still lower than the ordered
moment reported for \lmo.\cite{moussa} A part of the moment reduction may be attributed to the two-dimensional
character of the antiferromagnetic coupling in \lsmoee, but this reduction gives only $\Delta m =0.22$\mB.
However, there is evidence that part of the magnetic order in \lsmoee{} remains two-dimensional in character and
does not transform into the 3-dimensional Bragg intensities.\cite{senff-unpub}

The crystal and magnetic structure of the \lsmoacht -sample was studied by powder neutron diffraction. We find
the same magnetic superstructure peaks and obtain an ordered magnetic moment of 2.4(1)\mB. The further reduction
of the moment compared to pure \lsmoee{} agrees with the reduced N\'eel temperature, $T_N\sim$62\ K, and the
finding that, for this composition, the amount of two-dimensional diffuse magnetic scattering is further
enhanced.\cite{senff-unpub}

The most interesting structural aspect concerns the elongation of the MnO$_6$ octahedron. Three-valent Mn
possesses four electrons in the 3d-shell with one occupying the $e_g$-orbitals; therefore Mn$^{3+}$ is strongly
Jahn-Teller active, and one expects a strong octahedron elongation. In the layered material there is a
competition between the crystal field of the Mn planar structure favoring an elongation along the $c$ axis, like
in La$_2$CuO$_4$ or in La$_2$NiO$_4$ (note that the octahedron in the nickelates is elongated though Ni$^{2+}$ is
not a Jahn-Teller ion), and the exchange energy.\cite{khomskii} In order to allow for the virtual hopping of
$e_g$-electrons to their nearest neighbors in the $a,b$-plane, it is preferable to have  $e_g$-orbitals occupied
with the lobes within this plane. Furthermore, the elastic interactions will favor an arrangement such that the
long axis of one octahedron is pointing towards the short axis of an neighboring site.\cite{khomskii03} A
similar arrangement is realized in the perovskite \lmo{} in the $a,b$ planes of the Pbnm crystal
structure,\cite{carvajal} where it causes a ferromagnetic coupling.\cite{moussa} In the layered manganate \lsmoee
, the crystal field appears to overrule these mechanisms, since the elongation axes point along the $c$
direction. The Mn-O bond distances at room temperature amount to $2\times2.267\text{\ \AA}+4\times1.893\text{\
\AA}$ compared to $2\times1.968\text{\ \AA}+2\times1.907\text{\ \AA}+2\times2.178\text{\ \AA}$ for the perovskite
\lmo.\cite{carvajal} The difference of the Mn-O distances is even stronger in the layered material than in \lmo \
due to the combined effect of the Jahn-Teller effect and the \knif -structure crystal field. The ferroorbital
ordering of the $e_g$-orbital along the $c$ direction agrees with the antiferromagnetic ordering of the spins
within the planes.

\begin{figure}[tp]
\resizebox{0.95\figwidth}{!}{
\includegraphics*{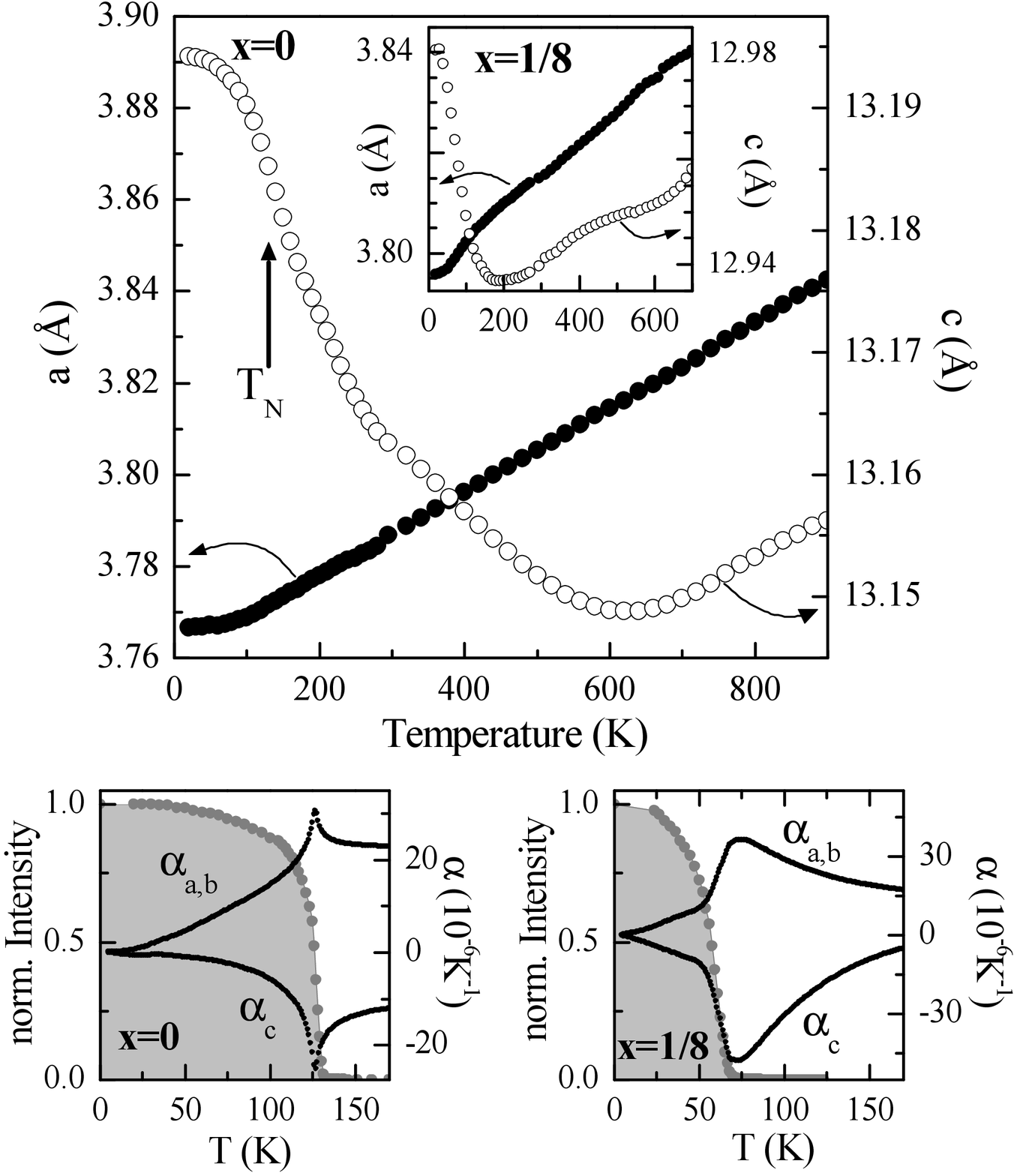}
} \caption{ (Upper Panel) Temperature dependence of the lattice
constants in \lsmoee{} as a function of temperature (the arrow
indicates the N\'eel-temperature). The inset shows the temperature
dependence of the lattice constants in \lsmoacht{} where N\'eel
order occurs at $\sim62$\ K. (Lower Panel) thermal expansion
coefficient (parallel and perpendicular to the MnO$_2$ layers)
determined by capacitance dilatometry and the normalized
intensity of the antiferromagnetic superstructure reflection (0.5
0.5 0) measured by neutron diffraction for \lsmoee{} (left) and
for \lsmoacht{} (right).} \label{fig3}
\end{figure}

In view of the large splitting in the bond distances, one might expect a complete orbital ordering and a stable
crystal structure. However, we find that the bond distances vary significantly with temperature (see Table I).
The Mn-O2 to Mn-O1 distance ratio increases from 1.197(2) to 1.209(3), upon cooling from room temperature to 20\
K. Further insight is obtained from the analysis of the lattice constants as a function of temperature, see Fig
3. Both, the determination of the lattice parameters by x-ray powder diffraction and the measurement of the
thermal expansion coefficient with a high resolution capacitance dilatometer, yield large and anisotropic
anomalies around the N\'eel ordering. Upon cooling, the $c$ axis expands while the in-plane parameters shrink, in
agreement with the observation of the more elongated MnO$_6$-octahedron at low temperature. The thermal expansion
coefficients, corresponding to the temperature derivatives of the lattice constants, exhibit a $\lambda$-like
anomaly at the N\'eel-temperature. In the antiferromagnetically ordered phase, the $c$ axis expansion and the
in-plane shrinking are less pronounced. The clear coupling between the structural effect and the magnetic
ordering indicates an orbital origin. Most likely the orbital order is not complete in \lsmoee, but some of the
in-plane $e_g$-orbitals are still occupied. The anisotropic thermal expansion indicates that there is a change in
the orbital occupation. Upon cooling, less in-plane orbitals are occupied. The temperature dependence of the
lattice parameters suggests that the orbital rearrangement is spread over a wide temperature interval, 10--600\
K. Only at very high temperatures does the $c$ axis exhibit a normal, positive thermal expansion.

In Fig. 2 the $e_g$ orbitals relevant in \lsmoee{} are depicted. The mixing of in-plane and out-of-plane
$e_g$-orbitals has been modeled by a combination of the $3z^2-r^2$ and $x^2-y^2$
orbitals,\cite{khomskii,nagai}
\[
   |\theta\rangle =cos(\theta/2)|3z^2-r^2\rangle+sin(\theta/2)|x^2-y^2\rangle .
\]
The $x^2-y^2$ and $3z^2-r^2$ orbitals correspond to the coefficients $\theta =\pi$  and $\theta = 0$,
respectively. In-plane elongated $e_g$-orbitals correspond to values of $\theta=\pm{2\pi}/3$. With such a linear
combination one assumes at least a local distortion of the tetragonal symmetry. The value of $\theta$ may
fluctuate around the average value of $\theta$=0 corresponding to the $3z^2-r^2$-orbital. However, at the moment
there is no evidence for such symmetry reduction. The orbital occupation in \lsmoee{} may eventually correspond
to a linear combination with imaginary coefficients corresponding to an admixture of orbitals which would not
break the local symmetry.\cite{Brink01}

The second sample presenting an antiferromagnetic ordering, \lsmoacht, exhibits qualitatively similar anomalies
in the lattice constants. However, the anomalous $c$ axis expansion and the in-plane shrinking occur over a
smaller temperature interval upon cooling. The $c$ parameter of \lsmoacht{} increases by the same amount between
200 and 10\ K, as does the $c$ axis in \lsmoee{} between 600 and 10\ K. Concerning the effects in the in-plane
parameters, it is less obvious that one may separate the electronically-induced effects from the normal thermal
expansion, as they have the same sign. The thermal expansion temperature dependence, see lower part of Fig. 3,
again shows the maxima just at the N\'eel temperature and a reduced effect in the ordered phase. So, both
compositions exhibit a continuous elongation of the MnO$_6$ octahedron when approaching the N\'eel order from
high temperature and this tendency gets stopped in the ordered state.

Doping \lsmoee{} with additional Sr induces Mn$^{4+}$-sites where no $e_g$-orbitals are occupied. Such
Mn$^{4+}$-sites may attract the $e_g$-orbitals of neighboring Mn$^{3+}$-site, and the concept of an orbital
polaron has been proposed.\cite{brink} In the layered material the doping-induced Mn$^{4+}$-sites will
destabilize the ordering of the $e_g$-orbitals elongated along the $c$ direction and -- simultaneously -- the
antiferromagnetic order. The balance between the crystal field preferring a $c$ elongation of the $e_g$-orbitals
and the kinetic energy favoring an orientation of the orbitals along the planes, is shifted towards the in-plane
orientation through the doping, since, only in this case, the $e_g$-electron may easily hop. Therefore, compared
to \lsmoee , there seems to be more in-plane elongated $e_g$-states occupied in \lsmoacht{} at high temperature.
The increase of the magnetic correlations induced upon cooling leads then to a redistribution of electrons which
is stronger than in \lsmoee.

The interpretation of the structural anomalies by an orbital redistribution is supported by recent x-ray
absorption measurements which find a finite and temperature-dependent occupation of the $e_g$-orbital oriented in
the planes.\cite{merz-unp}

The anomalies in the thermal expansion suggesting a rearrangement of the orbital occupation induced by the
magnetic interaction resemble recent observations on LaTiO$_3$ \cite{hemberger,cwik} and on
Ca$_2$RuO$_4$.\cite{braden1998} In these compounds, one finds an enhancement of an octahedron distortion upon
approach of the N\'eel order. In the case of the ruthenate the accompanying change in the orbital occupation has
been directly observed by x-ray absorption spectroscopy.\cite{hao} As in \lsmoee , the octahedron distortion
implies a stronger crystal field splitting of the orbital energies in these $t_{2g}$ systems. Observing a similar
effect in the strongly Jahn-Teller active $e_g$-system \lsmoee{} appears more astonishing.

\begin{figure}[tp]
\resizebox{0.95\figwidth}{!}{
\includegraphics*{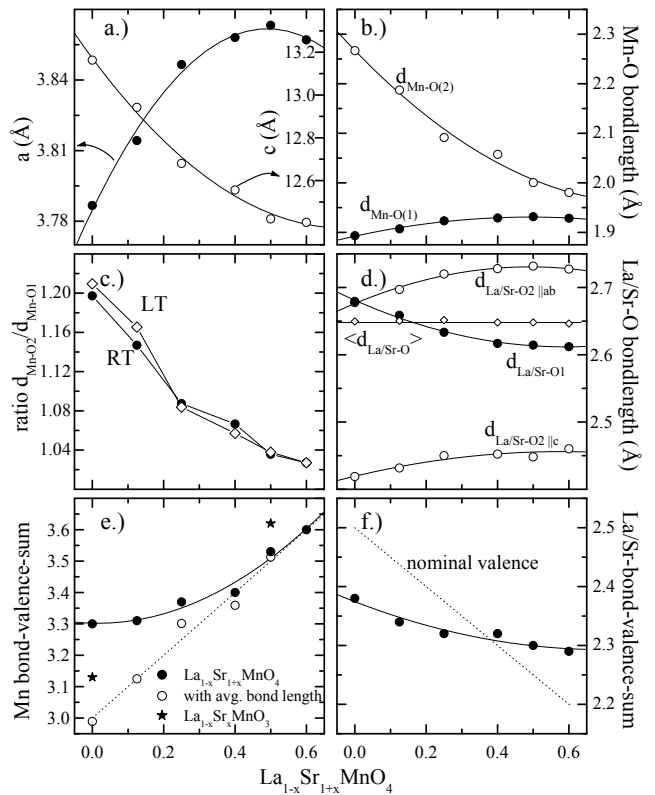}
} \caption{ Doping dependence of : the lattice parameters (a), the
Mn-O bond lengths  (b), the ratio of the
two Mn-O bond lengths (c), the three distinct bond
lengths in the (La,Sr)O$_9$-polyedron (d) for the series
La$_{1-x}$Sr$_{1+x}$MnO$_4$. Calculated bond-valence-sum for the
central Mn-ion (closed circles denote the values obtained with
the split bond lengths, open circle those calculated with the average Mn-O bond length and stars
values calculated for \lmo \cite{carvajal} and La$_{0.5}$Sr$_{0.5}$MnO$_3$ \cite{radaelli} e)) and for the La/Sr place (f). Drawn lines are
guides to the eye.}\label{fig4}
\end{figure}

\section{Average crystal structure for intermediate doping}

--{\it Crystal structure}-- Upon further increase of the Sr content the commensurate antiferromagnetic order, as
illustrated in Fig. 2, disappears, and rather complex schemes of charge, orbital and magnetic order appear,
amongst which that of the half-doped composition is best
studied.\cite{sternlieb,murakami,larochelle,wilkins,dhesi} The single crystal of composition \lsmoviert{} studied
here does not show the superstructure peaks indexed by ($h/2$ $k/2$ l) but magnetic scattering at an
incommensurate \vq -position (0.5,0.16,0), which one may not relate with the simple commensurate magnetic
ordering.\cite{senff-unpub} With the diffraction techniques used here, we are not able to study the structural
distortion arising from the charge and orbital order, we may only discuss the average crystal structure. A single
crystal of \lsmoef \ was studied on the four-circle diffractometer. As extinction effects were again severe, we
used the same methods as for the \lsmoee -sample. The data sets of 1030(554) reflections taken at room
temperature (T=25\ K) are described with reliability factors of $R_w(F^2)$=5.65\%(6.24\%) and
$R_{unw.}(F^2)$=5.65\%(6.07\%). The structural results are given in Table I.

The lattice constants, see Fig. 4a), show that the trend already discussed continues with a further increase of
the Sr content up to $x=0.6$. The $c$ parameter shrinks and the in-plane parameter slightly expands.
Simultaneously, the Mn-O2 bond shrinks and the Mn-O1-bond expands (see Fig. 4b) and 4c)). The increase of the Sr
content corresponds to an oxidation of the Mn-sites or to an increase in the number of Mn$^{4+}$-sites, which, in
first view, should lead to shorter average Mn-O bond distances. The Sr dependence of the lattice constants and
bond lengths indicates that this effect superposes with the one discussed above : the continuing change in the
orbital occupation at the Mn$^{3+}$-sites. For $x=0$ mainly the $c$-elongated orbitals are occupied, whereas, for
$x=0.5$, there is a strong occupation of the in-plane components. The reminiscent elongation of the MnO$_6$
octahedra along the c-direction at high Sr-concentrations can be understood due to the remaining influence of the
planar crystal structure. Note that the non-Jahn Teller configuration $3d^8$ in La$_2$NiO$_4$ still exhibits an
octahedron elongation due to the purely structural effects.\cite{struc-lanio} The temperature dependence of the
bond distance ratio (see Fig. 4c)), shows that the enhancement of the octahedron elongation at low temperature
occurs only in the low-doped compounds which exhibit the commensurate antiferromagnetic order.

The change of the Mn-O bond distances may be analyzed quantitatively using the bond valence sum formalism :
\begin{equation*}
 V_{Mn} =\sum_{i}exp\left({d_0 - d_i(Mn-O)\over B}\right), \\
\end{equation*}
where $d_0=(1-x)\cdot1.76\text{\AA}+x\cdot1.753\text{\AA{}}$ and $B$=0.37\AA{} are empirical
parameters\cite{brown} and $d_i(Mn-O)$ denotes the six Mn-O bond distances. The Mn bond valence sums are shown in
Fig. 4e); only for Sr content higher than $x=0.4$ one does find the nominal valence,
suggesting that for low
doping the crystal structure exhibits an internal stress similar to the
isostructural cuprates.\cite{struc-lacuo}
The internal stress seems to be released with further doping as the calculated bond valence sum approaches the nominal value. In Fig. 4e) we also show the bond-valence sums caluclated with the
averaged Mn-O distances, which follow the nominal values. It seems that part of the internal
stress is related to the strong splitting of the Mn-O bond distances, which is probably unsufficiently decribed with the bond valence scenario.
For comparison we also show the bond valence sums obtained for the perovskites La$_{1-x}$Sr$_x$MnO$_3$
which are quite close to the nominal values for x=0 and 0.5.\cite{carvajal,radaelli}

The La-O coordination polyhedron, too, changes significantly with Sr content, in particular the shortest La-O2
bond, parallel to the $c$ axis, which increases rapidly with doping in the range $0\leq x<0.4$, see Fig. 4d). The
analysis of the bond valence sums calculated via :
\begin{eqnarray*}
V_{La/Sr} =(0.5-0.5x)\times\sum_{i}{exp\left({d_{0-La} - d_i(La/Sr-O)\over B}\right)} \nonumber \\
+ (0.5+0.5x)\times\sum_{i}exp{\left({d_{0-Sr} - d_i(La/Sr-O)\over
B}\right),}
\end{eqnarray*}
is shown in Fig. 4f). At low doping, the obtained values are significantly lower than the expected values,
confirming the interpretation that there is an internal strain in this material. This strain is reduced with
doping and even turns into the opposite as the bond valence sums are larger than the expected values in the
samples with Sr concentration of 0.4 or higher.

--{\it Temperature dependence of the lattice parameters}-- The temperature dependencies of the lattice constants
for large doping are shown in Fig. 5. In contrast to the anomalous behavior found for the low-doped samples, the
thermal expansion is quite normal in this Sr-concentration range. The complex charge, orbital and magnetic
ordering schemes occurring in these samples does not imply structural changes similar to those induced through
the commensurate antiferromagnetic order.

\begin{figure}[tp]
\resizebox{0.95\figwidth}{!}{
\includegraphics*{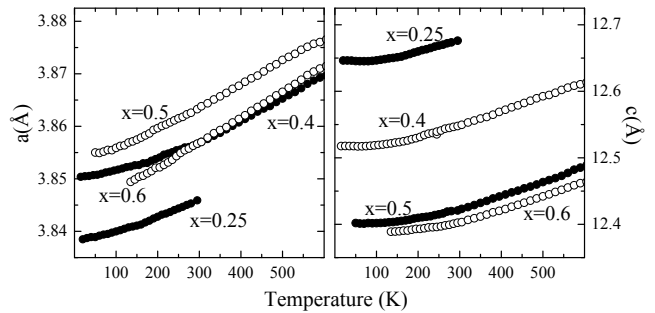}
} \caption{Temperature dependence of the lattice parameters in \lsmo ~ for concentrations $x>0.25$.}\label{fig5}
\end{figure}

--{\it Debye Waller factors} -- Our interpretation of the doping-induced reorientation of the $e_g$-orbitals is
well supported by the anisotropic atomic displacement parameters (ADP). Due to the intrinsic disorder induced by
the occupation of the same site by La and Sr, some of the displacement parameters are significantly higher than
one would expect from the phonon contributions. The disorder effect has been studied in detail for \lsco , where,
however, only 8\% of the La are replaced by Sr.\cite{braden2001} In \lsmo{} the doping disorder is maximal for
\lsmoee{} and decreases with further doping. Since the (La/Sr)-O bonds are perpendicular to the Mn-O bonds, the
doping disorder will displace the O-atoms mainly perpendicular to the Mn-O bonds. The U$_{11}$ parameter of O2
and the U$_{33}$ parameter of O1 are most sensitive. These displacement parameters are indeed maximal for $x=0$
and decrease with increasing Sr-content. The inter-atomic interaction potentials should be similar to those in
\lsco, therefore one may compare to the additional ADP obtained for the cuprate-system.\cite{braden2001}
Extrapolating the disorder induced ADP contributions calculated for \lsco \ to x=1.0 one obtains
$\Delta$U$_{11}$(O2)=0.023\AA $^2$ and $\Delta$U$_{33}$(O1)=0.0047\AA$^2$ and these values are of the order of
the low temperature ADP's measured in \lsmoee . The manganate seems to be less sensitive to the disorder most
likely due to the more stable character of its crystal structure. In the \lsmo{} phase diagram there are no
structural phase transitions, whereas \lsco{} exhibits the octahedron tilt instabilities.

\begin{figure}[tp]
\resizebox{0.95\figwidth}{!}{
\includegraphics*{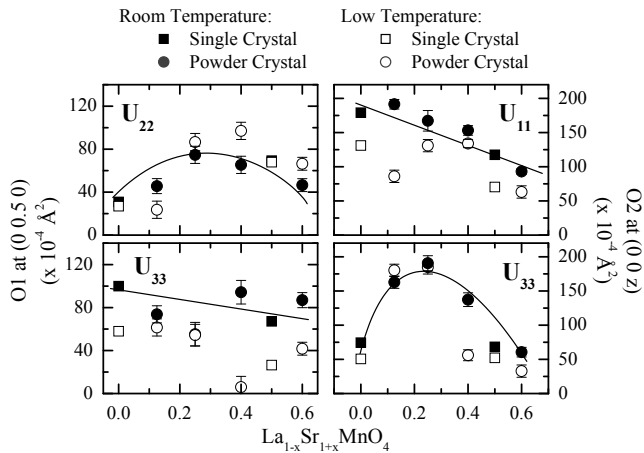}
} \caption{ Anisotropic displacement parameters of the two oxygen sites as function of temperature and doping.
Note that U$_{22}$(O1) and U$_{33}$(O2) correspond to the displacement of the oxygen parallel to the bonds.
}\label{fig6}
\end{figure}
The orbital occupation is related to the atomic displacement parameters parallel to the bonds, U$_{33}$ of O2 and
the U$_{22}$ parameter of O1. These parameters are smaller in pure \lsmoee{} than in intermediate
Sr-concentrations, although the doping disorder is maximal for that composition. U$_{33}$ of O2 exhibits a
maximum around $x=0.25$ where the orbital orientation is the less defined. This concentration roughly represents
the border between the majority occupation of $c$-axis elongated orbitals around \lsmoee \ and the majority
occupation of in-plane elongated orbitals at higher doping.

The longitudinal displacement parameter of the in-plane oxygen is maximal near half-doping.\cite{note}  This
finding agrees well with the model of the CE-type ordering of both the Mn$^{3+}$/Mn$^{4+}$-sites and orientation
of the $e_g$-orbitals.\cite{sternlieb,murakami,larochelle} Both displacements have not been taken into account in
our refinement of the average structure. The doping dependence of the U$_{22}$-O1 parameter suggests local
displacements of the order of 0.1\AA{} near half doping.

\section{Conclusions}

The crystal and magnetic structure of \lsmo{} was studied by several diffraction techniques and high resolution
capacitance dilatometry. At low doping $0.0\leq x\leq0.25$, the MnO$_6$ octahedra are elongated along the $c$
direction in accordance with a majority occupation of the $3z^2-r^2$ orbital. However, pronounced and anisotropic
anomalies in the thermal expansion yield evidence that the orbital occupation varies with temperature. Upon
cooling there seems to occur a shift of electrons towards the $3z^2-r^2$ orbitals. Upon increase of the
Sr-concentration, the amount of Mn$^{3+}$ ions is reduced and the octahedron elongation along the $c$-direction
diminishes. This effect has to be attributed to a majority occupation of the in-plane elongated $e_g$-orbitals,
as it was predicted to arise from the magnetic coupling between Mn$^{3+}$ and Mn$^{4+}$ sites.\cite{brink}

{\bf Acknowledgments} This work was supported by the Deutsche Forschungsgemeinschaft through the
Sonderforschungsbereich 608. We are grateful to D. Khomskii for interesting discussions.


\begin{thebibliography}{28}
\expandafter\ifx\csname
natexlab\endcsname\relax\def\natexlab#1{#1}\fi
\expandafter\ifx\csname bibnamefont\endcsname\relax
  \def\bibnamefont#1{#1}\fi
\expandafter\ifx\csname bibfnamefont\endcsname\relax
  \def\bibfnamefont#1{#1}\fi
\expandafter\ifx\csname citenamefont\endcsname\relax
  \def\citenamefont#1{#1}\fi
\expandafter\ifx\csname url\endcsname\relax
  \def\url#1{\texttt{#1}}\fi
\expandafter\ifx\csname
urlprefix\endcsname\relax\def\urlprefix{URL }\fi
\providecommand{\bibinfo}[2]{#2}
\providecommand{\eprint}[2][]{\url{#2}}


\bibitem{mang-gen} R. von Helmolt, J. Wecker, B. Holzapfel, L. Schultz and K.
Samwer, Phys. Rev. Lett. {\bf 71}, 2331 (1993); S. Jin, M. McCormack, T. H. Tiefel and R. Ramesh, J. Appl. Phys.
{\bf 76}, 6929 (1994); Y. Tokura, A. Urushibara, Y. Morimoto, T. Arima, A. Asamitsu, G. Kido and N. Furukawa, J.
Phys. Soc. Jpn. {\bf 63}, 3931 (1994).

\bibitem{magres-327} Y. Moritomo, A. Asamitsu, H. Kuwahara and Y. Tokura, Nature
(London) {\bf 380}, 141 (1996).

\bibitem{mag-214}M. Tokunaga, N. Miura, Y. Moritomo and Y. Tokura, Phys. Rev. B
{\bf 59}, 11151 (1999).

\bibitem{sternlieb}B. J. Sternlieb, J. P. Hill, U. C. Wildgruber, G. M. Luke,
B. Nachumi, Y. Moritomo and Y. Tokura, Phys. Rev. Lett. {\bf 76}, 2169 (1996).

\bibitem{murakami} Y. Murakami, H. Kawada, H. Kawata, M. Tanaka,
T. Arima, Y. Moritomo and Y. Tokura, Phys. Rev. Lett. {\bf 80}, 1932 (1998).

\bibitem{wilkins} B. Wilkins, P. D. Spencer, P. D. Hatton, S. P. Collins,
M. D. Roper, D. Prabhakaran and A. T. Boothroyd, Phys. Rev. Lett. {\bf 91}, 167205 (2003).

\bibitem{dhesi}S. S. Dhesi, A. Mirone, C. De Nada\"{i}, P. Ohresser, P. Bencok,
N. B. Brookes, P. Reutler and A. Revcolevschi, Phys. Rev. Lett. {\bf 92}, 056403 (2004).

\bibitem{larochelle} S. Larochelle, A. Mehta, N. Kaneko, P. K. Mang, A. F.
Panchula, L. Zhou, J. Arthur and M. Greven, Phys. Rev. Lett. {\bf 87}, 095502 (2001).

\bibitem{kawano} S. Kawano, N. Achiwa, N.Kamegashira and M. Aoki, J. de Physique {\bf 49}(C8), 829 (1988).

\bibitem{moussa} F. Moussa, M. Hennion, J. Rodr\'{i}guez-Carvajal, H. Moudden, L. Pinsard and A. Revcolevschi, Phys.
Rev. B {\bf 54}, 15149 (1996).

\bibitem{larger-m} M. Bieringer and J. E. Greedan, J. Mater Chem. {\bf 12}, 279
(2002).

\bibitem{bao} W. Bao, C. Chen, S. Carter and S.-W. Cheong, Sol. State Commun. {\bf 98}, 55 (1995).

\bibitem{moritomo} Y. Moritomo, Y. Tomioka, A. Asamitsu, Y. Tokura and Y. Matsiui, Phys. Rev. B {\bf 51}, 3297
(1995).

\bibitem{reutler} P. Reutler, O.
Friedt, B. B\"uchner, M. Braden and A. Revcolevschi, Journal of Crystal Growth {\bf 249}, 222 (2003).

\bibitem{baumann} C. Baumann, G. Allodi, B. B\"uchner, R. De Renzi, P. Reutler
and A. Revcolevschi, Physica B {\bf 326}, 505 (2003).

\bibitem{fullprof} J. Rodr\'{i}guez-Carvajal, Physica B {\bf 192}, 55(1993).

\bibitem{braden-cgo} M. Braden, G. Wilkendorf, J. Lorenzana, M. Ain, G.J. McIntyre, M. Behruzi, G. Heger, G. Dhalenne and
A. Revcolevschi, Phys. Rev. B {\bf 54}, 1105 (1996).

\bibitem{senff-unpub} D. Senff, Diploma-thesis, unpublished.

\bibitem{khomskii} D. I. Khomskii and G. Sawatzky, Sol. State
Commun. {\bf 102}, 87 (1997); K. I. Kugel and D. I. Khomskii, Sov. Phys. Usp. {\bf 25}, 231 (1982).

\bibitem{khomskii03} D.I. Khomskii and K. I. Kugel, Phys. Rev. B {\bf 67},
134401 (2003).

\bibitem{carvajal} J. Rodr\'{i}guez-Carvajal, F. Moussa, M. Hennion, H. Moudden, L. Pinsard and A. Revcolevschi,
Phys. Rev. B {\bf 57}, 3189 (1998)



\bibitem{nagai} T. Nagai, T. Kimura, A. Yamazaki, Y. Tomioka, K. Kimoto, Y. Tokura and Y. Matsui, Phys. Rev. B {\bf 68}, 092405 (2003).

\bibitem{Brink01} J. van den Brink and D. Khomskii, Phys. Rev. B {\bf 63}, 140416(R) (2001)

\bibitem{brink}G. Khilian and G. Khaliullin, Phys. Rev. B {\bf 60}, 13458
(1999); T. Mizokawa, D. I. Khomskii and G. Sawatzky, Phys. Rev. B {\bf 63}, 24403 (2000).

\bibitem{merz-unp} M. Merz, unpublished results.

\bibitem{hemberger} J. Hemberger, H.-A. Krug von Nidda, V. Fritsch, J. Deisenhofer, S. Lobina, T. Rudolf, P. Lunkenheimer,
F. Lichtenberg, A. Loidl, D. Bruns, and B. B\"uchner, Phys. Rev. Lett. {\bf 91}, 066403 (2003).

\bibitem{cwik} M. Cwik, T. Lorenz, J. Baier, R. M\"uller, G. Andr\'e, F. Bour\'ee, F. Lichtenberg, A. Freimuth, R. Schmitz,
E. M\"uller-Hartmann and M. Braden, Phys. Rev. B {\bf 68}, 060401(R) (2003).

\bibitem{braden1998}M. Braden, G. Andr\'e, S.
Nakatsuji and Y. Maeno, Phys. Rev. B {\bf 58}, 847 (1998); O. Friedt, M. Braden, G. Andr\'e, P. Adelmann, S.
Nakatsuji and Y. Maeno, Phys. Rev. B {\bf 63}, 174432(2001).

\bibitem{hao}T. Mizokawa, L. H. Tjeng, G. A. Sawatzky, G. Ghiringhelli, O. Tjernberg, N. B. Brookes, H. Fukazawa, S. Nakatsuji and Y. Maeno
Phys. Rev. Lett. {\bf 87}, 077202 (2001).

\bibitem{radaelli}P. G. Radaelli,  D. E. Cox, M. Marezio and S-W. Cheong, Phys. Rev. B {\bf 55}, 3015(1997).


\bibitem{struc-lanio} J. D. Jorgensen, B. Dabrowski, S. Pei, D. R. Richards and D. G. Hinks
Phys. Rev. B {\bf 40}, 2187 (1989).


\bibitem{brown} I.D. Brown and D. Altermatt, Acta Crystal. B {\bf 41}, 244 (1985).


\bibitem{struc-lacuo} M. Braden, P. Schweiss, G. Heger, W. Reichardt, Z. Fisk,
K. Gamayunov, I. Tanaka and H. Kojima, Physica C {\bf 223}, 396 (1994).




\bibitem{braden2001} M. Braden,
M. Meven,  W. Reichardt,  L. Pintschovius, M. T. Fernandez-Diaz, G. Heger, F. Nakamura and T. Fujita, Phys. Rev.
B {\bf 63}, 140510 (2001).

\bibitem{note} This conclusion is in particular supported by the neutron
single crystal data, which  yield more reliable ADPs. Additional studies
by x-ray single crystal diffraction \cite{merz-unp} confirm the ADP's qualitatively.


\end{thebibliography}
\end{document}